\documentclass{osa-article}

\journal{oe}

\usepackage{float}
\usepackage{braket} 
\usepackage{subfigure}
\usepackage{soul}

\articletype{Research Article}

\begin{document}

\title{Bandwidth control of the biphoton wavefunction exploiting spatio-temporal correlations}

\author{J.J. Miguel Varga,\authormark{1,2,*} Jon Lasa-Alonso,\authormark{1,2} Martín Molezuelas-Ferreras,\authormark{1} Nora Tischler,\authormark{4,5} and Gabriel Molina-Terriza\authormark{1,2,3,**}}

\address{\authormark{1}Centro de F\'isica de Materiales, Paseo Manuel de Lardizabal 5, 20018 Donostia-San Sebasti\'an, Spain.\\
\authormark{2}Donostia International Physics Center, Paseo Manuel de Lardizabal 4, 20018 Donostia-San Sebasti\'an, Spain.\\
\authormark{3}IKERBASQUE, Basque Foundation for Science, Mar\'ia D\'iaz de Haro 3, 48013 Bilbao, Spain.\\
\authormark{4}Centre for Quantum Computation and Communication Technology (Australian Research Council), Centre for Quantum Dynamics, Griffith University, Brisbane, QLD 4111, Australia.\\
\authormark{5}Dahlem Center for Complex Quantum Systems, Freie Universität Berlin, 14195 Berlin, Germany.}

\email{\authormark{*}miguelvarga@gmail.com}
\email{\authormark{**}gabriel.molina.terriza@gmail.com} 



\begin{abstract}
In this work we study the spatio-temporal correlations of photons produced by spontaneous parametric down conversion. In particular, we study how the waists of the detection and pump beams impact on the spectral bandwidth of the photons. Our results indicate that this parameter is greatly affected by the spatial properties of the detection beam, while not as much by the pump beam. This allows for a simple experimental implementation to control the bandwidth of the biphoton spectra, which only entails modifying the optical configuration to collect the photons. Moreover, we have performed Hong-Ou-Mandel interferometry measurements that also provide the phase of the biphoton wavefunction, and thereby its temporal shape. We explain all these results with a toy model derived under certain approximations, which accurately recovers most of the interesting experimental details.
\end{abstract}

\section{Introduction}

Photons are a natural choice for quantum information protocols. Their importance lies in their ease of transmission, the simplicity of their handling and their versatility regarding the degrees of freedom that can be used to encode information. Nowadays, the use of light in experimental studies of quantum phenomena range from their use in quantum cryptography protocols~\cite{Sit:s}, to exploring the propagation of entanglement in biological tissues~\cite{Shi2016}. They are also very useful in fundamental studies such as testing quantum foundations~\cite{Shadbolt2014} or exploring light-matter interactions in the nanoscale~\cite{PhysRevLett.121.173901, Lasa_Alonso_2020}.

One particular technical advance that has allowed this wide range of applications of photons in the quantum regime is the production of photonic states in nonlinear processes, such as spontaneous parametric down conversion (SPDC). In bulk optics SPDC, a strong pump beam is focused in a nonlinear crystal. This process produces pairs of quantum correlated photons with, typically, a very low efficiency. The versatility of this technique resides in the fact that the generated pairs of photons can be manipulated in several degrees of freedom (DOFs). In particular, we find a plethora of works that study entanglement of photons in polarization~\cite{PhysRevA.60.R773}, frequency~\cite{PhysRevLett.103.253601}, orbital angular momentum~\cite{Torres:04} or linear transverse momentum~\cite{PhysRevLett.94.100501}. Usually, the study of entanglement properties implies considering only one DOF of the photons. This can be done either by filtering and post selecting specific values of the other DOFs, or engineering them for minimal entanglement. However, it is also possible to study the properties of photons that are entangled in two or more DOFs, as in the case of hyperentangled photons~\cite{PhysRevLett.95.260501, Graham2015, PhysRevLett.122.123607}.

In particular, the correlations between spatial and temporal (or equivalently, spectral) DOFs are specially important in SPDC processes~\cite{Osorio_2008}. The question of how these variables influence each other has been present in the field for several decades. For example, in Ref.~\cite{PhysRevA.50.3349} correlations in the spatial coordinates of the generated biphotons were studied in terms of the pump's spectral bandwidth. The ability of tuning the spectro-temporal distribution is also fundamental for several applications. For example, obtaining information about the electronic structure of a medium strongly depends on the spectral shape of the photons~\cite{de_J_Le_n_Montiel_2013}. Furthermore, shaping the spectral distribution of the photons allows information processing techniques, such as orthogonal spectral coding~\cite{PhysRevLett.112.133602}.

Several methods for shaping the spectro-temporal distribution of entangled photons have been reported in the literature. For instance, in Ref.~\cite{PhysRevA.67.053810} a method was proposed to generate entangled photons with controllable frequency correlations that consisted on initiating counter-propagating SPDC into a single-mode nonlinear waveguide with a pulsed pump beam. In Ref.~\cite{PhysRevLett.99.243601} it was demonstrated that it is possible to perform a tunable control of the spectra by shaping the pump beam spatially. More sophisticated methods involve combining a quantum light pulse  with a spectrally tailored classical field~\cite{PhysRevLett.106.130501} or, directly customizing the nonlinear crystal's poling period~\cite{PhysRevA.93.013801}.

The different methods that have been used to control the biphoton wavefunction are variations of the following techniques: 1) Direct manipulation of the photons with external modulators or photon interference~\cite{Lima:09, Kagalwala2017, PhysRevLett.123.143601}, 2) customization of the nonlinear crystal characteristics and poling~\cite{Branczyk:11, PhysRevApplied.8.024035}, 3) tuning the temporal shape of a pulsed pump beam~\cite{Lu:18, Ansari:18} or 4) tailoring the spatial distribution of a monochromatic pump field~\cite{PhysRevA.73.063802, Romero_2012}. In this work we demonstrate that, when pumping a nonlinear crystal with a monochromatic Gaussian beam, it is possible to tune the spectral features of SPDC biphotons, typically measured by Hong-Ou-Mandel (HOM) interferometry~\cite{PhysRevLett.59.2044}, by changing the widht of the detection beam waist. This technique exploits the spatio-temporal correlations of the photons generated in the crystal. Our proposal is versatile and its experimental implementation is straightforward, as it only encompasses an adequate choice of collimation and coupling lenses. Under standard experimental conditions, the spectral bandwidth of the biphoton wavefunction can be tuned from $0.25$ nm to $1.25$ nm, i.e. by a factor of five. Interestingly, we find that this behavior cannot be reproduced by changing the pump beam width. We experimentally show that the spectral bandwidth of the photons depends very weakly on the waist of the Gaussian pump beam and we have developed an analytical model which accurately reproduces the main features of this phenomenon.

This paper is organized as follows: in Sec.~\ref{sec:exp_res} we describe the experimental setup used to measure the HOM interference maps and show the results obtained for different pump and detection beam waists. In Sec.~\ref{sec:disc} we explain the functional dependence of the biphoton spectrum shape on both the pump and detection beam waist. We also analyze how these parameters affect the temporal amplitude of the biphoton wavefunction. Finally, in Sec.~\ref{sec:concl} we sum up the main conclusions of the work.

\section{Experiment}\label{sec:exp_res}

\begin{figure}[H]
\centering\includegraphics[width=\textwidth]{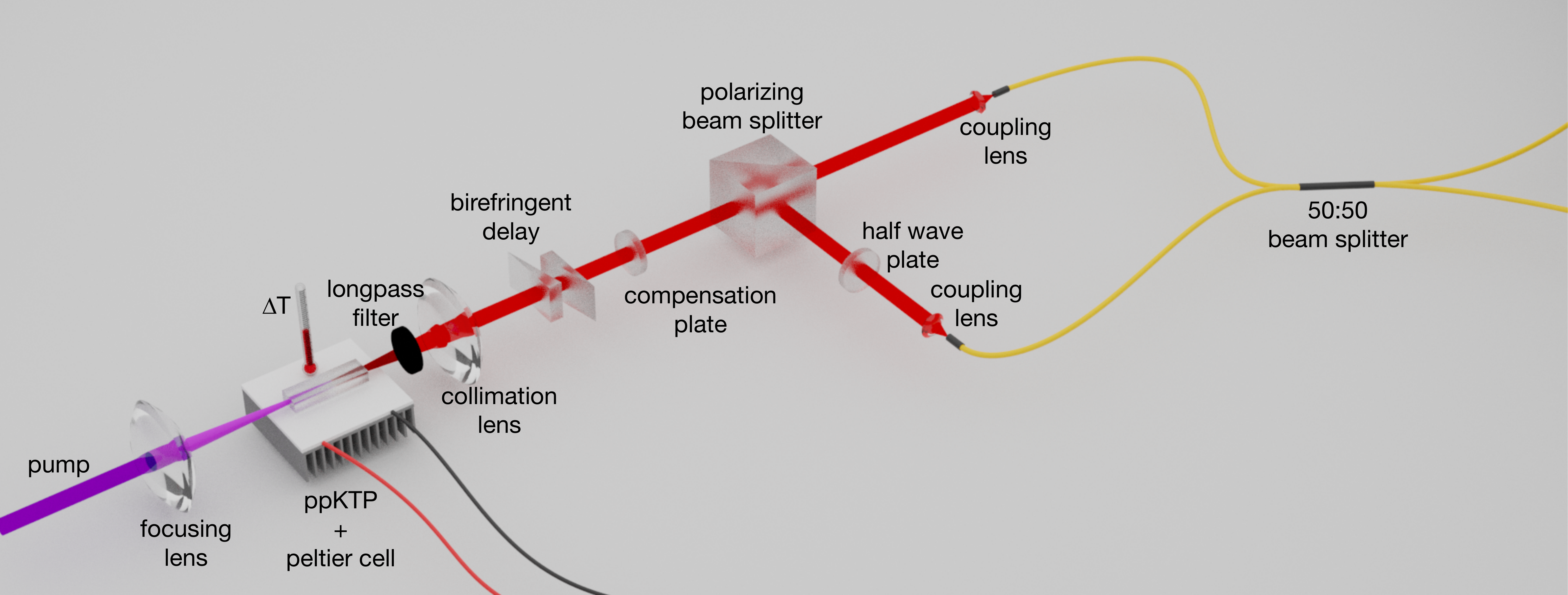}
\caption{Experimental setup: a monochromatic pump beam centered at $405.5~\text{nm}$ and with a power of $42~\text{mW}$ is focused onto a temperature-controlled $15~\text{mm}$ long ppKTP crystal. The down-converted light passes through a longpass filter and it is subsequently collimated by a lens. The delay of the photons is controlled by a birefringent compensator made by two calcite wedges and a calcite plate. Then, the photon pairs are separated by a polarizing beam splitter and focused into a polarization maintaining fiber beam splitter. A half-wave plate is used in one of the two arms in order to maximize the interference. After interference, a coincidence detection is performed by means of two avalanche photodiodes, one for each fiber beam splitter output arm.}
\label{fig:figure1}
\end{figure}

The experimental setup used in the measurement of the HOM maps is shown in Fig.~\ref{fig:figure1}. Photon pairs are generated by pumping a periodically poled potassium titanyl phosphate (ppKTP) crystal in a collinear, type-II down-conversion configuration. After the photons leave the nonlinear crystal, their temporal delay is modified by a set of birefringent plates: on the one hand, there is a compensation crystal which accounts for the temporal walk-off of the photons in the crystal and, on the other hand, we use a birefringent compensator consisting on two calcite wedges and a calcite plate. By moving the wedges laterally, the compensator allows us to perform a temporal delay among the photons up to $6.10~\text{ps}$. The photons are first separated with a polarizing beam splitter, subsequently coupled to single-mode fibers and finally recombined on the two ports of a fiber-based beam splitter after their polarization has been adjusted for maximum interference. The temperature of the crystal can be tuned and is actively stabilized by Peltier cells and a temperature controller with a precision of $0.1^{\circ}\text{C}$. At $T_0=54^{\circ}\text{C}$ the photons are both emitted at a central wavelength of $\lambda_d = 811~\text{nm}$ such that the emission can be considered to be degenerate. By varying the temperature, $T$, it is possible to tune the wavelengths, breaking this degeneracy~\cite{Fedrizzi:07}. Therefore, by changing the temperature, the photon wavelength is displaced. After interference, the photons are detected by an avalanche photodiode and their coincidences are quantified by means of a coincidence counting unit based on an FPGA board~\cite{Park:15}. 

In order to study the effect of the pump beam waist ($w_p$) and detection beam waist ($w_d$) on the biphoton spectrum, we changed the focusing and coupling lenses. In this way, it was possible to perform measurements for different configurations of pump and detection modes given by $w_p$ and $w_d$. For each configuration of modes, we proceeded to measure a HOM map. We used the technique developed in~\cite{PhysRevLett.115.193602}, where on top of temporally delaying the photons prior to interfering them, we also changed the temperature of the crystal. The typical temporal delay used in HOM interference only provides partial information about the spectral amplitude of the biphoton. However, in a continuous wave (cw) pumped generation of biphoton states, changing the temperature of the crystal and performing measurements of HOM dips for different temperatures provides full information about both the amplitude and phase of the biphoton wavefunction, allowing to extract the temporal correlations. In a nutshell, this is possible because the temperature scan amounts to a sort of autocorrelation in the frequency space. For more details about the technique and its limitations, please refer to~\cite{PhysRevLett.115.193602}. In our experiment, the measurements were taken by varying the temperature by $0.2^{\circ}\text{C}$ between $45^{\circ}\text{C}$ and $60^{\circ}\text{C}$ and the delay among the photons by $0.12~ \text{ps}$ between $-3.05~\text{ps}$ and $3.05~\text{ps}$. 

\begin{figure}[H]
\centering\includegraphics[width=\textwidth]{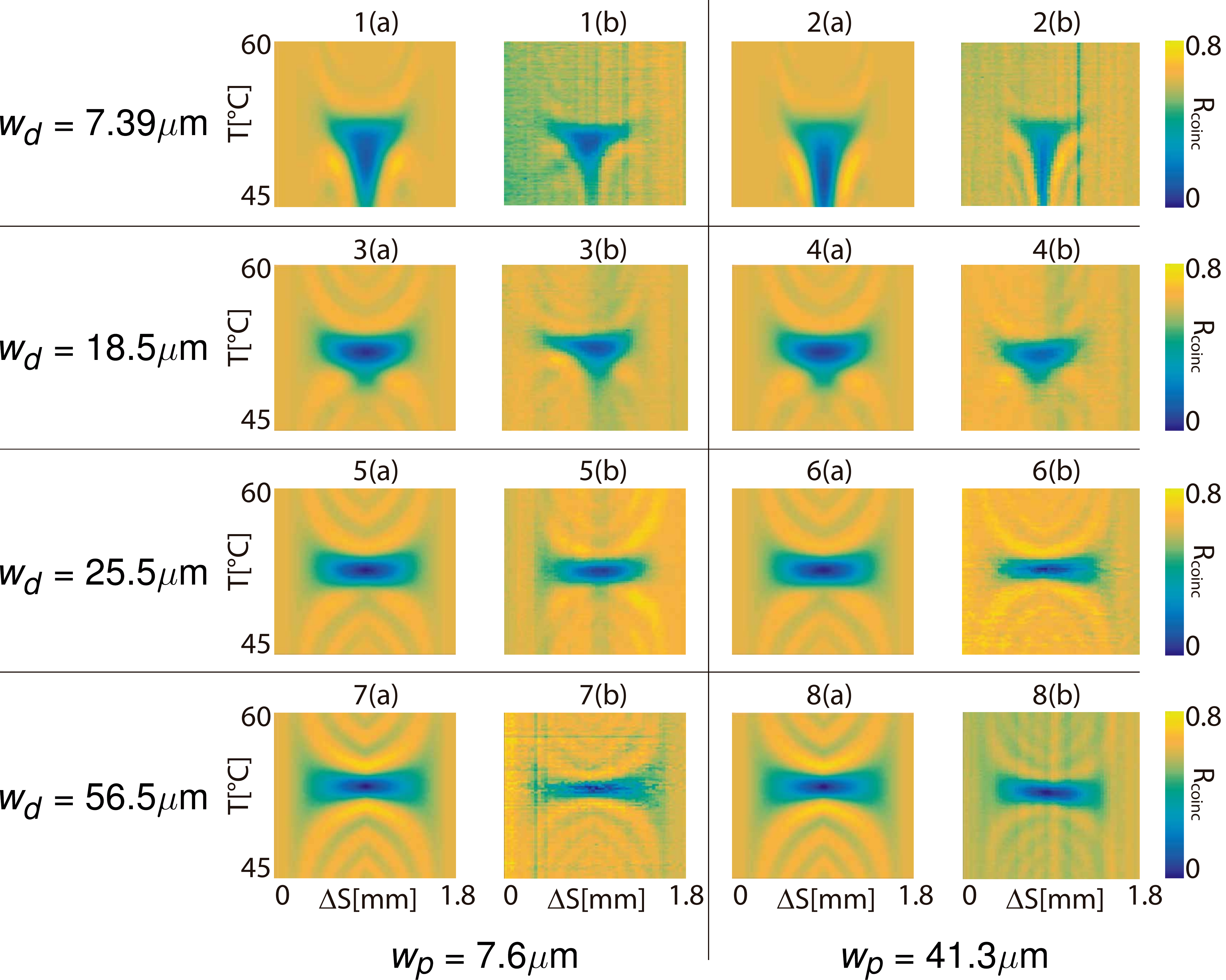}
\caption{Hong Ou Mandel maps for different pump beam waists $w_p$ and detection beam waists $w_d$. (a) are the theoretical and (b) the experimental results.  The maximum value of coincidences in $0.1$s for each case is (for maps from $1$ to $8$): $14899$, $3765$, $47754$, $61011$, $50118$, $71230$, $22438$ and $19071$.}   
\label{fig:figure2}
\end{figure}

The results are shown in Fig.~\ref{fig:figure2}. The rows correspond to HOM maps measured for a fixed detection beam waist $w_d$, whereas columns correspond to a given pump beam waist $w_p$. The columns labeled by an (a) correspond to the theoretical calculations and those labeled by a (b) to the experimental measurements. The color of the maps corresponds to the normalized number of coincidences per $0.1$s. In the case of the experimental maps we have subtracted the background corresponding to the accidental counts of the detector. The blue areas are the HOM dips. We can see that the agreement between the experimental and the theoretical results is excellent. From these results it becomes evident that the shape of the HOM maps, and hence the biphoton spectral wavefunction, depends on $w_d$ while it remains basically unaffected as $w_p$ is changed.

\section{Discussion}\label{sec:disc}

In order to explain the different behavior of the biphoton spectrum with respect to the beam waists $w_p$ and $w_d$, we have developed a toy model for the total mode function $\Phi(\omega_s,\omega_i,\mathbf{p},\mathbf{q})$, where $\mathbf{p} = (p_x,p_y)$ and $\mathbf{q}=(q_x,q_y)$ are the coordinates of the transversal momentum of the signal and idler photon, respectively. The full wavefunction contains the information of both the probability amplitude and the phase of the generated photons for different frequencies and momenta, i.e. $\ket{\Phi}=\int{\text{d}\omega_s \text{d}\omega_i \text{d}\mathbf{p} \text{d}\mathbf{q} \Phi(\omega_s,\omega_i,\mathbf{p},\mathbf{q}) a^\dagger_{\omega_s,\mathbf{p}} a^\dagger_{\omega_i,\mathbf{q}} \ket{0}}$. We start from the expression for the total mode function~\cite{PhysRevA.70.043817}:
\begin{equation}\label{eq:biphoton}
    \Phi(\omega_s,\omega_i,\mathbf{p},\mathbf{q})=\overline{E}_p(p_x+q_x,p_y+q_y)\textrm{sinc}(\Delta k_zL/2)\exp(-i\Delta k_z L/2),
\end{equation}
where $L$ is the crystal length, $\overline{E}_p$ is the amplitude of the pump beam in momentum space and
\begin{align}
    \Delta k_z &= k_p - k_s - k_i  - 2\pi/\Lambda
\end{align}
is the longitudinal phase matching condition. In the expression above $\Lambda=\Lambda(T)$ is the period of the poling of the crystal and we have defined
\begin{align}
    k_p &=\sqrt{(\omega^0_p n_p/c)^2 - (p_x+q_x)^2 - (p_y+q_y)^2}\\
    k_s&=\sqrt{(\omega_s n_s/c)^2 - p_x^2 - p_y^2}\\
    k_i&=\sqrt{(\omega_i n_i/c)^2 - q_x^2 - q_y^2}.
\end{align}
Here, $k_j$, $\omega_j$ and $n_j$ are the wavenumber, angular frequency and refractive index of the pump ($j=p$), signal ($j=s$) and idler ($j=i$) photons, respectively. $c$ is the speed of light in vacuum.

In our case, we are pumping the crystal with a continuous wave laser with a narrow bandwidth, which means that the frequencies of the signal and idler photons are linked by the energy conservation condition $\omega_p=\omega_s+\omega_i$.  In this situation, the only interesting parameter is the frequency difference of, say, the signal and the degenerate frequency: $\Omega=\omega_s-\omega_p/2=\frac{\omega_s - \omega_i}{2}$. Moreover, after coupling the photons to single mode fibers, we erase the spatial degrees of freedom. Thus, to obtain the biphoton spectral wavefunction, $\Phi(\Omega, T)$, we have to integrate Eq.~(\ref{eq:biphoton}) weighted by the Gaussian detection mode functions $G(\mathbf{p})$ and $G(\mathbf{q})$, with respect to $\mathbf{p}$ and $\mathbf{q}$:
\begin{equation}
    \Phi(\Omega, T) = \int d\mathbf{p} d\mathbf{q} \Phi(\Omega,\mathbf{p},\mathbf{q},T) G^\ast(\mathbf{p}) G^\ast(\mathbf{q}).
\end{equation}
At this point, we assume that we stand at the degeneracy temperature $T=T_0$ and make some approximations. We consider that $k_j\gg p_j, q_j$, that the refractive indices for the three photons (pump, signal and idler) are the same and that the $\textrm{sinc}$ function can be approximated by an exponential function with a factor in the exponent, i.e. $\textrm{sinc}(x)\sim\textrm{exp}(-0.455|x|)$~\cite{PhysRevA.53.4360}. Then, the biphoton spectral wavefunction reads as:
\begin{align}
    \Phi(\Omega) &= \frac{w_p w_d^2}{(2\pi)^{3/2}}\int d\mathbf{p} d\mathbf{q} \exp\left(-\frac{w_p^2((p_x+q_x)^2+(p_y+q_y)^2)}{4}\right)  \nonumber\\
    &\exp\left(-\frac{0.455L}{2} \left|F\Omega +\frac{1}{2k_{0p}(0)}\left[(p_x-q_x)^2+(p_y-q_y)^2\right]\right|\right)\nonumber\\
     &\exp\left(-i
     \frac{L}{2}\left[F\Omega +\frac{1}{2k_{0p}(0)}\left[(p_x-q_x)^2+(p_y-q_y)^2\right]\right]\right)\nonumber\\
     & \exp\left(-\frac{w_d^2(p_x^2+p_y^2+q_x^2+q_y^2)}{4}\right) \exp\left(i\frac{Lc}{2\omega_p n_{0p}}(p_x^2+p_y^2+q_x^2+q_y^2)\right),
\end{align}
where $k_{0p}=\frac{\omega_p n_{0p}}{c}$, $n_{0p}$ being the refractive index for the pump photon at degenerate temperature $T_0$, and $F \equiv - \frac{\partial k_s}{\partial \Omega}-\frac{\partial k_i}{\partial \Omega}$.

The absolute value of the second exponential imposes that the integral needs to be evaluated in two different regions, depending on whether $\Omega\geq0$ or $\Omega<0$. In both cases, the integral is analytic and results in: 
\begin{align}
    \Phi(\Omega\geq 0) &= A \exp\left(-\alpha \Omega\right), \label{eq:om_pos} \\
    \Phi(\Omega<0) &= A \left\{ \exp\left(-\alpha\Omega\right)\exp\left((w_d^2+w_{\text{opt}}^2) Fk_{0p}\Omega/4 \right)-\right.\nonumber\\
    & \left.-\frac{w_d^2+w_{\text{opt}}^2}{w_d^2-w_{\text{opt}}^2}\exp\left(\alpha^\ast\Omega\right)\left[\exp\left((w_d^2-w_{\text{opt}}^2) Fk_{0p}\Omega/4\right)-1\right]\right\}. \label{eq:om_neg}
\end{align}
Here, we have defined
\begin{align}
		A&=\frac{4 w_p w_d^2}{\sqrt{2\pi}(2 w_p^2+w_d^2-i w_{\text{opt}}^2/0.455 )(w_d^2+w_{\text{opt}}^2)},\\
    \alpha&=\frac{LF}{2}(0.455+i),\\
    w_{\text{opt}}^2&=2 \frac{0.455L c}{\omega_p n_{0p}}.
\end{align}

From Eqs.~(\ref{eq:om_pos}) and~(\ref{eq:om_neg}) we find that the pump waist $w_p$ only changes the overall amplitude of the biphoton spectral function, an effect that is not relevant since we normalize the spectra to have a unitary area under the curve. This normalization is considered because, experimentally, a change in the overall amplitude only implies a difference on the total amount of detected photons and, in practice, these differences are mainly caused by the alignment conditions present in each set of measurements. On the other hand, the detection beam waist $w_d$ affects the full functional form of the spectral wavefunction, not only the overall amplitude. Interestingly, we find that $w_d$ affects the form of $\Phi(\Omega)$ only for the negatives values of $\Omega$ and, hence, it does not only modify the spectral width but also breaks the symmetry of the wavefunction. The smaller the detection beam waist is, the greater the asymmetry of the spectral function becomes.

\begin{figure}[H]
\centering\includegraphics[width=0.6\textwidth]{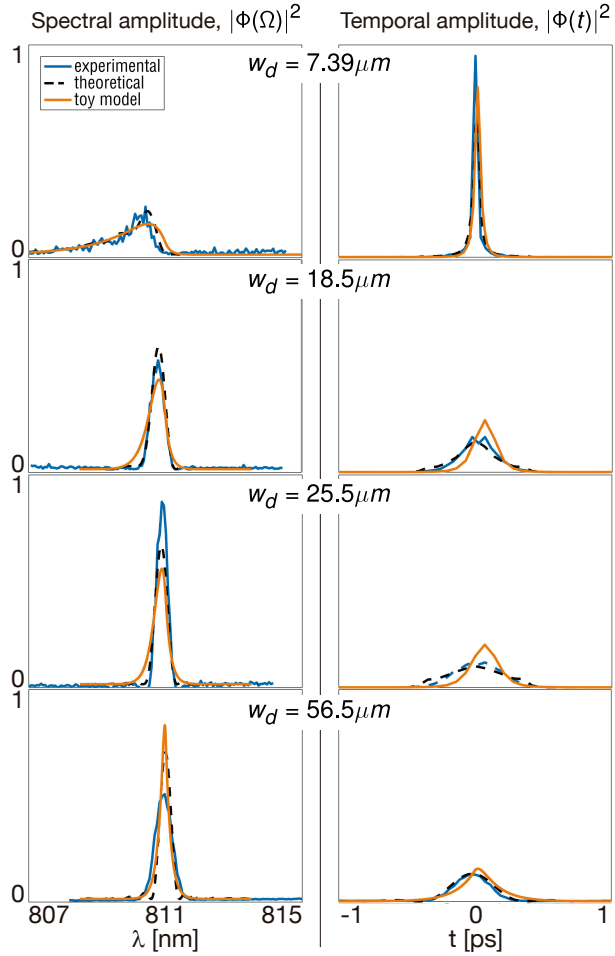}
\caption{Spectral and temporal amplitudes of the signal photon for different detection beam waists $w_d$. The pump beam waist is fixed to $w_p=7.6~\mu\text{m}$, corresponding to the HOM maps of the left column in Fig.~\ref{fig:figure2}. The amplitudes are normalized so that their area under the curve is unitary.}
\label{fig:figure3}
\end{figure}

This behavior is depicted in Fig.~\ref{fig:figure3}, where we show the biphoton spectral amplitude, $|\Phi(\Omega)|^2$, for different detection beam waists. We compare our toy model with the experimental and theoretical results obtained from the HOM maps in the first two columns of Fig.~\ref{fig:figure2}, that is, for a fixed pump beam waist $w_p=7.6~\mu\text{m}$. In the left column of Fig.~\ref{fig:figure3} the spectral amplitude obtained by means of the method described in Ref.~\cite{PhysRevLett.115.193602} is displayed. It can be checked that, as predicted by our toy model, when $w_d$ is decreased the wavefunction becomes asymmetric and wide. Contrarily, for larger values of $w_d$, the wavefunction becomes symmetric and narrow, converging its mean value to the degenerate wavelength, $\lambda_d=811~\text{nm}$. The temporal biphoton wavefunction, $\Phi(t)$, presented in the right column of Fig.~\ref{fig:figure3} is obtained by performing a discrete Fourier transform of the spectral function, $\Phi(\Omega)$. As expected, in contrast to the spectral function, the width of the temporal wavefunction becomes wider as we increase $w_d$. 

While in our model the pump beam width has no influence on the spectrum bandwidth, some of the experimental results (mainly Figs.~\ref{fig:figure2}-1 and \ref{fig:figure2}-2) indicate that this parameter slightly modifies the spectral shape of the wavefunction for smaller values of $w_d$. The reason why this is not described by our model is that we have assumed that the index of refraction is the same for the pump, signal and idler photons. This approximation imposes that the sinc function depends only on the difference of the momenta. Thus, the sinc function and the pump beam shape, $\overline{E}_p$, which depends on the sum of the momenta, are integrated independently. If the refractive indices are assumed not to be the same, this is no longer true and one can notice a small influence of the pump beam on the spectrum shape and bandwidth.

\begin{figure}[H]
\centering\includegraphics[width=\textwidth]{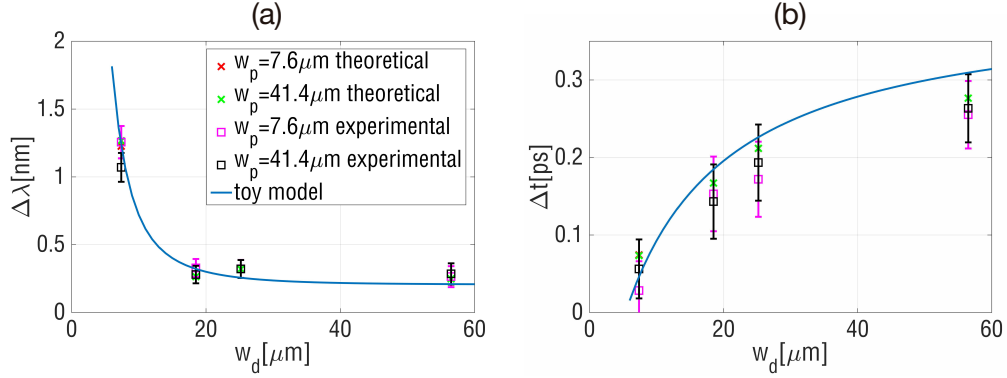}
\caption{Dependence of the width of the biphoton spectra on detection beam waist, $w_d$. We compare the results of the full numerical simulations (crosses), the experimental results (squares) and the toy model (solid line) for two values of the pump beam waist, $w_p$. (a) shows the wavelength width as a function of the detection beam waist. We see that the width of the biphoton spectra can be tuned by a factor of five. (b) shows the dependence of the temporal amplitude width on the detection beam waist. We find that the temporal width can also be experimentally tuned over a factor of five.}
\label{fig:figure4}
\end{figure}

The behavior of the photon's spectrum width is further clarified in Fig.~\ref{fig:figure4}, which shows both the dependence of the spectral and the temporal widths, $\Delta \lambda$ and $\Delta t$ respectively, on the detection beam waist. $\Delta\lambda$ and $\Delta t$ are computed as the standard deviations around the mean value of both the spectral and temporal amplitudes shown in Fig.~\ref{fig:figure3}. We have plotted $\Delta \lambda$ and $\Delta t$ comparing the widths obtained by means of theoretical calculations (crosses), experimental measurements (squares) and our toy model (solid line) for a set of waists that are realizable in a standard optics laboratory. From this figure we note that the wavelength width can be tuned within the interval $\Delta\lambda=[0.25,1.25]~\text{nm}$, namely, by a factor of five. On the other hand, we find that the temporal width can also be tuned by a factor of five approximately, that is, within the interval $\Delta t=[50,270]~\text{fs}$.

Briefly, the approximated expression of the biphoton wavefunction given in Eqs. \eqref{eq:om_pos} and \eqref{eq:om_neg} is able to accurately describe the behavior of biphoton spectra under usual experimental conditions. When comparing the results obtained by the toy model with the theoretical and experimental results, we find that the agreement is reasonably good, as it is shown in Fig.~\ref{fig:figure3} and Fig. \ref{fig:figure4}. Importantly, the approximated biphoton wavefunction allows us to predict the behavior of the spectral and temporal amplitude widths as a function of the detection beam waist $w_d$. As we have just shown, this is a crucial parameter in the design and characterization of twin photons sources based on bulk SPDC.

\section{Conclusion}\label{sec:concl}

To sum up, we have presented a simple and effective method to shape the biphoton spectro-temporal wavefunction. The technique is based on tailoring the spatial coupling of photons produced by SPDC into a single-mode fiber. To study the change in the spectro-temporal amplitude, we have first performed HOM interferometry measurements varying both the pump waist and the detection beam waist. To understand the behavior of the spectrum in terms of these two parameters, we have developed a toy model for the spectral biphoton wavefunction. We have analytically shown that $w_p$ affects only the global amplitude of the spectral function but not its functional form. On the other hand, the dependence of the spectral wavefunction shape on the detection beam waist is fundamental. Finally, we have found that this method allows tuning the width of the spectro-temporal amplitudes of the biphoton wavefunction by approximately a factor of five.

\section{Acknowledgements}

JJMV, JLA, MMF and GMT acknowledge the FIS2017-87363-P project of the Spanish Ministerio de Educación, Cultura y Deporte. NT acknowledges support by the Griffith University Postdoctoral Fellowship Scheme and by the Alexander von Humboldt Foundation. This article is dedicated to the memory of Juan José Sáenz.

\section{Disclosures}

The authors declare no conflicts of interest.

\bibliography{sample}

\end{document}